\documentclass[universe,communication,accept,moreauthors,pdftex]{mdpi} 

\firstpage{1} 
\pubvolume{1}
\issuenum{1}
\articlenumber{0}
\pubyear{2021}
\copyrightyear{2021}
\externaleditor{Academic Editor: {Włodzimierz Piechocki }} 
\datereceived{28 October 2021} 
\dateaccepted{07 December 2021} 
\datepublished{} 
\hreflink{https://doi.org/} 

\Title{{Singularities in Inflationary Cosmological Models}}

\TitleCitation{Singularities in Inflationary Cosmological Models}

\Author{Leonardo Fern\'andez-Jambrina \orcidA{}}

\AuthorNames{L. Fern\'andez-Jambrina}

\AuthorCitation{Fern\'andez-Jambrina, L.}

\address[1]{Matem\'atica e Inform\'atica Aplicadas,
E.T.S.I. Navales, Universidad Polit\'ecnica de Madrid, 
Avenida de la Memoria 4, E-28040 Madrid, Spain; leonardo.fernandez@upm.es}

\abstract{Due to the accelerated expansion of the universe, the
possibilities for the formation of singularities has changed from the
classical Big Bang and Big Crunch singularities to include a number of
new scenarios.  In recent papers it has been
shown that such singularities may appear in inflationary cosmological
models with a fractional power scalar field potential.  In this paper
we enlarge the analysis of singularities in scalar field cosmological
models by the use of generalised power expansions of their Hubble
scalars and their scalar fields in order to describe all possible models
leading to a singularity, finding other possible cases.  Unless a
negative scalar field potential is considered, all singularities are
weak and of type IV.}


\begin{document}
%

\section{Introduction}

The formation of singularities is one of the major challenges for
general relativity and has lead to plenty of conjectures and results
\cite{HE} in the field of mathematical relativity.  Even the term
singularity comprises many different definitions, such as curvature
singularities, geodesic incompleteness and b-incompleteness. 
Concerning cosmological scenarios, until the end of the 20th
century, the only possibilities of formation of singularities were 
the initial Big Bang singularity and, in the case of spatially closed 
cosmological models, the final Big Crunch singularity, since energy 
conditions were enforced in these cosmological models.

However, the discovery that our universe is undergoing accelerated expansion, even \mbox{presently 
~\cite{snpion,perlmutter,Davis:2007na,WoodVasey:2007jb,Leibundgut:2004,wmap,wmap1}},
triggered the search for either new ingredients for the energy content
of the universe (dark energy)
\cite{Padmanabhan:2006ag,Sahni:2006pa} or for 
 modifications of the general theory of gravity that could 
cope with such a feature 
\cite{Maartens:2007,Durrer:2007re,modigravi}.

Since these modifications of the standard cosmological model have usually 
involved violation of at least one energy condition, the main 
theorems on singularities were not applicable and new types of singular 
behavior were added to the list, which was enlarged from the original 
proposal \cite{Nojiri:2005sx} several times \cite{IV,yurov,sesto}.  
An updated list may be found in \cite{grandrip,ptrsa} in terms of the scale 
factor $a(t)$, the Hubble scalar $H(t)$ and the equation of state of
the model:

\begin{itemize}    
   \item Type $-$1. ``Grand bang/rip''. Ref. \cite{grandrip} The scale factor
   becomes null or diverges for $w=-1$.  

   \item Type 0. ``Big bang''. The scale factor becomes null for 
   $w\neq-1$.  
   
   \item Type I. ``Big rip'' \cite{Caldwell:2003vq}. The scale 
   factor diverges for $w\neq-1$.  

   \item Type II. ``Sudden singularities'' 
   \cite{sudden,odin,sudden1,lake,odin1,mariusz,luis,mariusz1,barrow2,barrow3,odin2,barrow4,barrow5,denk,suddenfirst,suddenferlaz},
   also named ``quiescent singularities'' \cite{quiescent}. The
   derivatives of the scale factor diverge from second derivative on.
   Some special cases are named "big brake" \cite{brake} and "big boost"
   \cite{boost}. 

   \item Type III. ``Big freeze'' \cite{freeze} or ``finite scale factor 
   singularities''. The
   derivatives of the scale factor diverge from first derivative on.
   
   \item Type IV. ``Generalised sudden singularities \cite{tsagas}. 
   The derivatives of the scale factor diverge from a derivative
   higher than second onward.
   
 \item Type V. ``$w$-singularities'' \cite{wsing,loitering}. No 
 derivatives of the scale factor diverge, but the 
 barotropic index $w=p/\rho$ in the equation of state does \cite{barotrope}.
   
\item Type $\infty$. ``Directional singularities'' 
\cite{hidden,initial}.
These happen at an infinite value of the coordinate time, but they can
be reached in finite proper time by only some observers.  In \cite{HE}
they are dubbed p.p. curvature singularities (curvature singularities
along a parallelly transported basis) in a more general framework.
\end{itemize}

Some of these singularities ($-$1, 0, I, $\infty$) are strong 
\cite{ellis,tipler,krolak,rudnicki,puiseux} in the sense that they distort finite objects, whereas 
others are weak (II,IV,V) and cannot be considered either the beginning or the 
end of the universe.

On the other hand, such singular behavior has been shown not to appear 
just as possible initial or final stages of the universe, but also in 
inflationary models \cite{graham,graham1}.

In that paper the authors consider cosmological models endowed with a scalar 
field $\phi$ under a potential $V(\phi)=A \phi^{n}$, with $A>0$, $n>0$. 
These models have been thoroughly studied for the integer $n$, since they 
provide classical examples of inflationary models for even $n$. 
However, if we allow $n$ to be a non-integer, a possibility that has been 
considered in the literature \cite{fractional,fractional1}, new features appear in the 
form of sudden, generalised singularities of type IV:
\begin{itemize}
%
	
	\item If $k<n<k+1$, where $k$ is a natural number, the derivatives
	of the Hubble scalar diverge from the $(k + 2)$th onward.
\end{itemize}

These singularities are not singularities in the scalar polynomials of
the curvature, but in their derivatives and are named $C^k$ scalar
curvature singularities in \cite{ellis1}. They were considered the 
first examples of realistic cosmological models wherein these 
singularities appeared.

In this paper we would like to enlarge the analysis of \cite{ellis1}
by considering other scalar field potentials, making use of
generalised power expansions of the Hubble scalar, the scalar field
and the potential in the cosmic time coordinate.  In
Section~\ref{hubble} we perform the power expansion of the Hubble
parameter and the scalar field and derive the singularities that may
appear.  Since the derivative of the Hubble parameter is to be
negative, it is seen that just singularities of types II, IV ad V may
appear in these models.  In Section~\ref{scalar} we relate the
expansion of the Hubble parameter and the expansion of the scalar 
field in order to derive the form of the scalar field potential of 
the models as a power expansion in the scalar field $\phi$.  A final
section of conclusions is included, where we show the list of scalar
field models that develop singularities, which are seen to be of the 
weak type IV, unless a negative potential is considered, which allows 
type II singularities. The list of potentials leading to 
singularities includes and enlarges the one enclosed in 
\cite{graham,graham1}.

\section{Divergences in the Hubble Parameter and the Scale Factor\label{hubble}}

We focus on spatially flat homogeneous and isotropic spacetimes
endowed with a metric \begin{equation}ds^2=-dt^2+a^2(t)\left(dr^2+
r^2\left(d\theta^2+\sin^2\theta
d\phi^2\right)\right),\label{metric}\end{equation} denoting by $a(t)$,
the scale factor of the universe, using the cosmological time $t$ as
the time coordinate.  If we write down Einstein equations for these
spacetimes, due to their symmetry, we end up with just two equations,
\begin{equation}\label{flrw}\rho= \frac{3\dot a^2}{a^2},\qquad
p=-\frac{2\ddot a}{a}-\frac{\dot a^2}{a^2},\end{equation} where
$\rho(t)$ is the energy density and $p(t)$ is the pressure of the
content of the universe.  Derivatives with respect to $t$ are denoted 
by a dot. The Hubble parameter, a measure of the expansion of the 
universe, is defined as $H=\dot a/a$.

According to this, the equation of state for the content of the 
universe, $p=w\rho$, can be written in terms of a coefficient $w$, 
\begin{equation}w=\frac{p}{\rho}=-\frac{1}{3}-\frac{2}{3}\frac{a\ddot a}{\dot a^2},\end{equation}
which has been dubbed the barotropic index, which is closely related to
the deceleration parameter $q$, \begin{equation}q=-\frac{a\ddot a}{\dot
a^2}=\frac{1+3w}{2},\end{equation} and the rest of cosmographical parameters can 
be obtained in a similar fashion \cite{cosmography}.

In the case of a scalar field $\phi$ subject to a potential $V(\phi)$,
 energy density splits into a kinetic energy and a potential energy 
term,
\begin{equation}\rho=\frac{\dot\phi^2}{2}+V(\phi),\end{equation}
and so does  pressure,
\begin{equation}p=\frac{\dot\phi^2}{2}-V(\phi).\end{equation}

The Friedman equations for this configuration are then \cite{liddle}
\begin{equation}\label{friedman}
3H^2=\frac{\dot\phi^2}{2}+V(\phi),\qquad
\dot H=-\frac{\dot\phi^2}{2}.\end{equation}

Deriving the first equation and replacing $\dot H$ with the help of 
the second equation, we get the evolution equation for the scalar 
field,
\begin{equation}\ddot \phi+3H\dot\phi+V'(\phi)=0.\end{equation}

We expect a slow-rolling scenario, for which the potential $V(\phi)$
dominates over the kinetic term $\dot\phi^2$ in the first Friedman
equation.  If this is true, it requires a positive scalar field
potential.

The use of generalised power expansions in cosmological time has been shown useful for 
analysing features such as singularities in cosmology \cite{visser}. 
Following this idea, we assume a generalised power expansion for the Hubble parameter
$H(t)$ around a time $t_{0}$,
\begin{equation}H(t)=h_{0}(t_{0}-t)^{\xi_{0}}+h_{1}(t_{0}-t)^{\xi_{1}}+h_{2}(t_{0}-t)^{\xi_{2}}+\cdots,\end{equation}
with $\xi_{0}<\xi_{1}<\cdots$. In order to have expansion, we require 
$h_{0}>0$.

Since, according to the second Friedman Equation (\ref{friedman}), the derivative of the Hubble parameter, \begin{equation}\dot 
H(t)=-\xi_{0}h_{0}(t_{0}-t)^{\xi_{0}-1}-\xi_{1}h_{1}(t_{0}-t)^{\xi_{1}-1}
-\xi_{2}h_{2}(t_{0}-t)^{\xi_{2}-1}-\cdots,\end{equation}
must be negative in order 
to allow a scalar field interpretation,
the product $\xi_{0}h_{0}$ must be positive. This restricts us to 
non-negative values for all the exponents $\xi_{i}$; and, if $\xi_{0}=0$, 
then $\xi_{1}h_{1}$ and hence $h_{1}$ must be positive.

We can relate $H(t)=\dot a(t)/a(t)$ to the scale factor $a(t)$ in order to check for 
singularities at $t_{0}$. Since
\begin{equation}\ln a(t)= \ln a_{0} 
-\frac{h_{0}}{\xi_{0}+1}(t_{0}-t)^{\xi_{0}+1}-
\frac{h_{1}}{\xi_{1}+1}(t_{0}-t)^{\xi_{1}+1}-\cdots\end{equation}
\begin{equation}a(t)=a_{0}e^{-\frac{h_{0}}{\xi_{0}+1}(t_{0}-t)^{\xi_{0}+1}-
\frac{h_{1}}{\xi_{1}+1}(t_{0}-t)^{\xi_{1}+1}-\cdots},\end{equation}
in terms of an integration constant $a_{0}=a(t_{0})$.

This leads to several possibilities:

\begin{itemize}
%
%
	
	\item $\xi_{0}=0$: We require positive $h_{0}$ and $h_{1}$. 
	We have  finite $a(t_{0})$ and $H(t_{0})$ and we 
 	have type II ($\xi_{1}<1$) or IV $(\xi_{1}\ge1$) singularities at $t_{0}$.

	\item $\xi_{0}>0$: The exponent $\xi_{0}+1$ and $h_{0}$ are
	positive. The scale factor $a(t_{0})$ is finite and $H(t_{0})$ 
	vanishes. Depending on the value of $\xi_{0}$ we 
	have singularities at $t_{0}$ of type II  ($\xi_{0}<1$) or of type IV ($\xi_{0}\ge1$) with a vanishing Hubble 
	factor. If $\xi_{0}\ge 2$ and every $\xi_{i}$ is 
	natural, we would have a type V singularity.
\end{itemize}
\section{\label{scalar}Reconstruction of the Scalar Field Potential}

We may reconstruct the scalar field potential $V(\phi)$ from the 
Friedman equations.
%

According to the second Friedman Equation (\ref{friedman}),
\begin{equation}\dot\phi(t)=\pm \sqrt{-2\dot H(t)}=\pm 
\sqrt{2h_{0}\xi_{0}}(t_{0}-t)^{(\xi_{0}-1)/2}+\cdots\end{equation}
the scalar field behaves as
\begin{equation}\phi(t)=\phi_{0}\mp\frac{2\sqrt{2h_{0}\xi_{0}}}{\xi_{0}+1}(t_{0}-t)^{(\xi_{0}+1)/2}+
\cdots,\end{equation}
in terms of an integration constant $\phi_{0}$, whereas we get the 
time evolution of the potential from the first Friedman equation,
\begin{eqnarray*}V(\phi(t))&=&3H^2(t)+\dot H(t)=
3h_{0}^2(t_{0}-t)^{2\xi_{0}}+6h_{0}h_{1}(t_{0}-t)^{\xi_{0}+\xi_{1}}
+3h_{1}^2(t_{0}-t)^{2\xi_{1}}
\\&+&6h_{0}h_{2}(t_{0}-t)^{\xi_{0}+\xi_{2}}+6h_{1}h_{2}(t_{0}-t)^{\xi_{1}+\xi_{2}}
+3h_{2}^2(t_{0}-t)^{2\xi_{2}}+\cdots
\\&-&\xi_{0}h_{0}(t_{0}-t)^{\xi_{0}-1}-\xi_{1}h_{1}(t_{0}-t)^{\xi_{1}-1}
-\xi_{2}h_{2}(t_{0}-t)^{\xi_{2}-1}-\cdots.
\end{eqnarray*}

For $\xi_{0}<-1$, the lowest order corresponds to the $2\xi_{0}$ term, 
whereas for $\xi_{0}>-1$ it is the $\xi_{0}-1$ one unless 
$\xi_{0}=0$, but we have already excluded models with negative 
$\xi_{0}$. However, negative $\xi_{0}$ has been considered for 
intermediate inflation models \cite{intermediate}.

Additionally, according to the slow roll approximation, the $2\xi_{0}$ term
is expected to dominate over the $\xi_{0}-1$ term.  This would happen
only for models with non-positive $\xi_{0}$.  Hence the only
inflationary models with positive potential are those with
$\xi_{0}=0$.  Inflationary models with $\xi_{0}>0$ would have a
negative scalar field potential.

Let us consider these cases:

\begin{itemize}

\item Non-natural $\xi_{0}>0$: $V(\phi(t))\simeq
-\xi_{0}h_{0}(t_{0}-t)^{\xi_{0}-1}\propto
\left(\phi(t)-\phi_{0}\right)^{\eta}\propto(t_{0}-t)^{(\xi_{0}+1)\eta/2}$,
with $\eta=2(\xi_{0}-1)/(\xi_{0}+1)$.  This means a leading power
$\eta\in(-2,2)$ in the negative potential and vanishing $H(t_{0})$.
For $\eta\in (-2,0)$ we have a type II singularity.  Otherwise we have
a type IV singularity.



%

\item $\xi_{0}=0$: This is the case of a finite $H(t_{0})$, 
\begin{equation}H(t)=h_{0}+h_{1}(t_{0}-t)^{\xi_{1}}+\cdots,\end{equation}\begin{equation}
\dot H(t)=-\xi_{1}h_{1}(t_{0}-t)^{\xi_{1}-1}+\cdots,\end{equation}\begin{equation}
\dot 
\phi(t)\simeq\pm\sqrt{2h_{1}\xi_{1}}(t_{0}-t)^{(\xi_{1}-1)/2},\end{equation} where 
$\xi_{1},h_{1}>0$. The expansions of the scalar field and the time 
evolution of the potential are, in this case,
\begin{equation}\phi(t)= 
\phi_{0}\mp\frac{2\sqrt{2h_{1}\xi_{1}}}{\xi_{1}+1}(t_{0}-t)^{(\xi_{1}+1)/2}+
\cdots,\end{equation}
\begin{eqnarray*}V(\phi(t))&=&
3h_{0}^2+6h_{0}h_{1}(t_{0}-t)^{\xi_{1}}
+3h_{1}^2(t_{0}-t)^{2\xi_{1}}
\\&+&6h_{0}h_{2}(t_{0}-t)^{\xi_{2}}+6h_{1}h_{2}(t_{0}-t)^{\xi_{1}+\xi_{2}}
+3h_{2}^2(t_{0}-t)^{2\xi_{2}}+\cdots
\\&-&\xi_{1}h_{1}(t_{0}-t)^{\xi_{1}-1}
-\xi_{2}h_{2}(t_{0}-t)^{\xi_{2}-1}-\cdots,
\end{eqnarray*}
\end{itemize}
leading to several interesting subcases:
\begin{itemize}
    \item  $\xi_{1}<1$: $V(\phi(t))\simeq 
    -\xi_{1}h_{1}(t_{0}-t)^{\xi_{1}-1}\propto 
    (\phi(t)-\phi_0)^{2(\xi_{1}-1)/(\xi_{1}+1)}$, and $V(\phi)$ behaves as 
    a power $(\phi-\phi_0)^{\eta}$ with $\eta\in(-2,0)$. For such models 
	$H(t_{0})$ is finite, but not $\dot H(t_{0})$. That is, a type II 
	singularity, but with a negative potential.
\end{itemize}
The rest of the subcases correspond to type IV singularities and 
positive scalar field potentials: 
\begin{itemize}
\item Non-natural $\xi_{1}>1$: $V(\phi(t))\simeq 3h_{0}^2
-\xi_{1}h_{1}(t_{0}-t)^{\xi_{1}-1}\simeq
3h_{0}^{2}-\alpha\xi_{1}h_{1}$ \linebreak$(\phi(t)-\phi_0)^{2(\xi_{1}-1)/(\xi_{1}+1)}$,
and the potential behaves as a positive constant plus a term
$(\phi-\phi_{0})^{\eta}$, with exponent
$\eta=2(\xi_{1}-1)/(\xi_{1}+1)$ in the interval $(0,2)$.  $\dot
H(t_{0})$ vanishes and $H^{(n)}(t_{0})$ diverges for
$\eta\in\left(\frac{2(n-2)}{n},\frac{2(n-1)}{n+1}\right)$, $n\ge2$.

	\item $\xi_{1}=1$: Type IV singularities with finite $\dot 
	H(t_{0})$ may arise in this case: 
\begin{equation}V(\phi(t))=3h_{0}^2-h_{1}+6h_{0}h_{1}(t_{0}-t)+3h_{1}^2(t_{0}-t)^{2}-\xi_{2}h_{2}(t_{0}-t)^{\xi_{2}-1}\cdots,\end{equation}
and a leading linear evolution for the scalar field,
\begin{equation}\phi(t)= \phi_0\mp\sqrt{2h_{1}}(t_{0}-t)+\cdots.\end{equation}

Again, several possibilities arise:
\begin{itemize}
	
\item $h_{1}\neq 3h_{0}^{2}$:

For $\xi_{2}\in(1,2)$, the leading non-constant term in $V(\phi(t))$ is the one with exponent $\xi_{2}-1$. The potential behaves as a
constant plus a term $(\phi-\phi_{0})^{\eta}$, with \mbox{$\eta=\xi_{2}-1\in(0,1)$}.  
The singularity appears for $\ddot H(t_{0})$.

For $\xi_{2}>2$ or $\xi_{2}=2$, $h_{2}\neq 3h_{0}h_{1}$, the leading non-constant term in $V(\phi(t))$ is the
linear one.  The potential behaves as a constant plus a linear term in
$\phi$.  The singularity would appear for $\dddot H(t_{0})$.


We may produce scalar field potentials with a higher leading power, 
besides the constant term, by requiring some cancellations between 
terms. We reproduce here the cases of quadratic, cubic and quartic 
powers:

For $\xi_{2}=2$, $h_{2}=3h_{0}h_{1}$,
\begin{equation}V(\phi(t))=3h_{0}^2-h_{1}+(18h_{0}^{2}+3h_{1})h_{1}(t_{0}-t)^{2}
-\xi_{3}h_{3}(t_{0}-t)^{\xi_{3}-1}+\cdots,\end{equation} and the potential behaves as a
constant plus a quadratic term in $\phi$.

\item $h_{1}= -6h_{0}^{2}$, $\xi_{2}=2$, $h_{2}=-18h_{0}^{3}$: 
\begin{equation}V(\phi(t))=9h_{0}^2-\xi_{3}h_{3}(t_{0}-t)^{\xi_{3}-1}
+648(t_{0}-t)^{3}+\cdots,\end{equation} 

For $\xi_{3}\in(2,4)$, the potential behaves as a constant plus a term
$\phi^{\eta}$, with $\eta=\xi_{3}-1\in(1,3)$.  The singularity appears
for $\dddot H(t_{0})$ if $\eta\in(1,2)$ and for $\ddddot H(t_{0})$ if 
$\eta\in(2,3)$.

For $\xi_{3}>4$, the potential behaves as a
constant plus a cubic term in $\phi$.  

For $\xi_{3}=4$, $h_{3}=162h_{0}^{5}$, $\xi_{4}\in(4,5)$,
\begin{equation}V(\phi(t))=9h_{0}^2-\xi_{4}h_{4}(t_{0}-t)^{\xi_{4}-1}+1944h_{0}^{6}(t_{0}-t)^{4}+\cdots,\end{equation} 
the potential behaves as a constant plus a term $(\phi-\phi_{0})^{\eta}$, with
$\eta=\xi_{4}-1\in(3,4)$.  $\ddddot H(t_{0})$ is regular now.

For $\xi_{3}=4$, $h_{3}=162h_{0}^{5}$, $\xi_{4}>5$:
\begin{equation}V(\phi(t))=9h_{0}^2-\xi_{4}h_{4}(t_{0}-t)^{\xi_{4}-1}+1944h_{0}^{6}(t_{0}-t)^{4}+\cdots,\end{equation}
and the potential behaves as a constant plus a quartic term in $\phi$.

We see clearly that they follow the singularity pattern in 
\cite{graham,graham1} for 
the derivatives of the Hubble parameter, despite the additional terms.

We may further remove the constant term in the scalar field potential 
to recover the cases in \cite{graham,graham1}. We produce some cases in order 
to show how the singular derivatives of the Hubble parameter come up:

	\item $h_{1}=3h_{0}^{2}$:
\begin{equation}V(\phi(t))=18h_{0}^{3}(t_{0}-t)+27 h_0^4 (t_{0}-t)^2-\xi_{2}h_{2}(t_{0}-t)^{\xi_{2}-1}\cdots\end{equation}

For $\xi_{2}\in(1,2)$, the potential goes as $(\phi-\phi_{0})^{\eta}$ with 
$\eta=\xi_{2}-1\in(0,1)$ and 
the singularity appears for $\ddot H(t_{0})$.

For $\xi_{2}>2$, the potential is linear in $\phi$. And the same 
happens for $\xi_{2}=2$, $h_{2}\neq 9h_{0}^{3}$. $\ddot H(t_{0})$ is 
regular now.

For $\xi_{2}=2$, $h_{2}= 9h_{0}^{3}$,
\begin{equation}V(\phi(t))=81h_0^4(t_{0}-t)^2+162h_{0}^5(t_{0}-t)^3-\xi_{3}h_{3}(t_{0}-t)^{\xi_{3}-1}+\cdots,\end{equation}
we have for $\xi_{3}\in(2,3)$ a potential which goes as $(\phi-\phi_{0})^{\eta}$ with 
$\eta=\xi_{3}-1\in(1,2)$ and a quadratic potential for $\xi_{3}>3$ or 
$\xi_{3}=3$, $h_{3}\neq 27h_{0}^{4}$. $\dddot H(t_{0})$ is 
regular in the latter case.

For $\xi_{2}=2$, $h_{2}= 9h_{0}^{3}$, $\xi_{3}=3$, $h_{3}= 
27h_{0}^{4}$,
\begin{equation}V(\phi(t))=324h_{0}^5(t_0-t)^3+729h_0^6(t_0-t)^4
-\xi_{4}h_{4}(t_{0}-t)^{\xi_{4}-1}+\cdots,\end{equation}
we have for $\xi_{4}\in(3,4)$ a potential which goes as $(\phi-\phi_{0})^{\eta}$ with 
$\eta=\xi_{4}-1\in(2,3)$ and a cubic potential for $\xi_{4}>4$ or 
$\xi_{4}=4$, $h_{4}\neq 81h_{0}^{5}$. $\ddddot H(t_{0})$ is 
regular in the latter case.

\end{itemize}
\end{itemize}

\section{Conclusions}

In the previous section we have identified the scalar field 
potentials that lead to singularities. We may collect them from the cases 
that have arisen.

For scalar field potentials of the form, $\eta_{0}<\eta_{1}<\cdots$,
\begin{equation}V(\phi)=V_{0}(\phi-\phi_{0})^{\eta_{0}}+V_{1}(\phi-\phi_{0})^{\eta_{1}}+\cdots\ldots,\end{equation}
we have obtained type IV singularities in
higher derivatives of the Hubble parameter:
\begin{itemize}
	
	\item $V(\phi)=V_{0}+\cdots+
	V_{n-2}(\phi-\phi_{0})^{n-2}+ V_{\eta}(\phi-\phi_{0})^{\eta}+
	\cdots $, $\eta\in (n-2,n-1)$, $n=2,3,\ldots$, and finite $\dot 
	H(t_{0})$: The first singular 
	derivative of the Hubble parameter is 
	$H^{(n)}(t_{0})$.  These include the cases studied in the paper
	by Barrow and Graham \cite{graham,graham1}.
	
	

	
	
	\item $V(\phi)=V_{0}+ V_{\eta}(\phi-\phi_{0})^{\eta}+
	\cdots,$ $\eta\in(0,2)$ and vanishing $\dot H(t_{0})$: $H^{(n)}(t_{0})$ is singular for 
	$\eta\in \left (\frac{2(n-2)}{n},\frac{2(n-1)}{n+1}\right)$, $n\ge2$.

%
\end{itemize}

Additionally, if we allow the scalar field potential to be negative:

\begin{itemize}
%
	
\item  $\eta_{0}\in (-2,0)$: Type II singularity.

\item  $\eta_{0}\in [0,2)$ and vanishing $H(t_{0})$: Type IV singularity.
\end{itemize}

Summarizing, we have performed a thorough analysis of the
singularities that may arise in scalar field cosmologies of the form
(\ref{friedman}) and obtained just Type IV singularities, except in the
case of a negative potential, for which Type II singularities may
appear.

These results extend, and of course comprise, those in
\cite{graham,graham1} for fractional power scalar field potentials.
Aside from the case of negative scalar field potentials and the extension
of the form of the potential with additional terms in natural powers
of $\phi$, we see that, on considering a vanishing value of $\dot
H(t_{0})$, singularities appear for potentials of the form
$V(\phi)=V_{0}+V_{\eta}(\phi-\phi_{0})^{\eta}+ \cdots,$
$\eta\in(0,2)$.  Such singularities may appear on higher derivatives
of the Hubble parameter, starting at $H^{(n)}(t_{0})$, for $n$ as large
as required.

These singularities are weak \cite{tipler,krolak,puiseux} and in this
sense they do not imply the end of the universe, since finite objects
may cross them without disruption and spacetime can be extended
beyond the singularity, which may be interpreted in terms of shock
waves \cite{cotsakis} propagating through spacetime.  Energy
conditions are not violated and energy density and pressure remain
finite and only higher derivatives diverge.

\vspace{6pt}

\funding{ This research received no external funding }

\acknowledgments{The author wishes to thank the University of the 
Basque Country for their hospitality.}

\conflictsofinterest{The author declares no conflict of interest  }
\end{paracol}
\reftitle{References}

\begin{thebibliography}{999}
	

\bibitem{HE}Hawking, S.W.; Ellis, G.F.R. 
\textit{The Large Scale Structure of Space-Time}; Cambridge University
Press: Cambridge, UK,  1973.

\bibitem{snpion}
Riess1, A.G; Filippenko, A.V.; Challis, P; Clocchiatti, A.;  Diercks, 
A; Garnavich, P.M; Gilliland, R.L.;  Hogan, C.J; Jha, S;  Kirshner 
R.P et al. Observational Evidence from Supernovae for an Accelerating Universe
and a Cosmological Constant.
\emph{Astron. J.} \textbf{1998}, \emph{116}, 1009.

\bibitem{perlmutter} Perlmutter, S.; Aldering, G.; Goldhaber, G.;
Knop, R. A.; Nugent, P.; Castro, P. G.; Deustua, S.; Fabbro, S.;
Goobar, A.; Groom, D. E. et al.  Measurements of $\Omega$ and 
$\Lambda$ from 42 
High-Redshift Supernovae.
\emph{Astrophys. J.} \textbf{1999}, \emph{517}, 565.
\bibitem{Davis:2007na}
Davis, T.M. ; M\"ortsell, E.; Sollerman, J; Becker,
A.C.; Blondin, S.; Challis, P.; Clocchiatti, A.;
Filippenko, A.V.; Foley, R.J.; Garnavich, P.M. et al.
Scrutinizing exotic cosmological models using {ESSENCE} supernova data
combined with other cosmological probes.  \emph{ Astrophys.  J.}
\textbf{2007}, {\em 666}, 716.


\bibitem{WoodVasey:2007jb}
Wood-Vasey, W.M.; Miknaitis, G.; Stubbs, C.W.; 
Jha, S.; Riess, A.G.; Garnavich, P.M.; Kirshner, R.P.; Aguilera, C.; 
Becker, A. C.; Blackman, J. W. \emph{et al.}
 Observational Constraints on the Nature of the Dark Energy: First
 Cosmological Results from the ESSENCE Supernova Survey. 
 \emph{Astrophys. J.} \textbf{2007}, {\em 666},  694.
\bibitem{Leibundgut:2004} Leibundgut, B. Cosmology with Supernovae. 
\emph{Reviews of Modern Astronomy}; Schielicke, R.E., Ed.; Wiley-VCH: Weinheim, Germany, 2004;  Volume {17}.
\bibitem{wmap}  Spergel, D.N.; Verde, L.; Peiris1, H.V; 
Komatsu, E.; Nolta, M.R.; Bennett, C.L.; Halpern, M.; Hinshaw, G.;  
Jarosik, N.; Kogut A. et al.  
 First Year Wilkinson Microwave Anisotropy Probe (WMAP)
Observations: Determination of Cosmological Parameters. 
 \emph{Astrophys. J. Suppl.} \textbf{2003}, {\em 148}, 175.
 
\bibitem{wmap1} Spergel, D.N.; Bean, R.; Dor\'e, O.; Nolta, M.R.; Bennett, C.L.; Dunkley, J.; Hinshaw, G.; Jarosik, N.; Komatsu, E.; Page, L.  et al. 
 Wilkinson Microwave Anisotropy Probe (WMAP) three year results:
Implications for cosmology.
 \emph{Astrophys. J. Suppl.} \textbf{2007},   {\em 170}, 377.
\bibitem{Padmanabhan:2006ag}
Padmanabhan, T. Dark Energy: Mystery of the Millennium. 
 \emph{AIP Conference Proceedings} \textbf{2006}, {\em 861}, 179.
\bibitem{Sahni:2006pa}
 Sahni, V.; Starobinsky, A. Reconstructing dark energy. 
 \emph{Int. J. Mod. Phys. D} \textbf{2006}, {\em 15}, 2105.
\bibitem{Maartens:2007} Maartens, R. Dark Energy and Dark Gravity. \emph{J. Phys. Conf. Ser.} \textbf{2007}, \emph{68},
012046.
\bibitem{Durrer:2007re}
 Durrer, R.; Maartens, R. Dark energy and dark gravity: theory 
 overview. 
\emph{Gen. Rel. Grav.}  \textbf{2008}, {\em 40}, 301.

\bibitem{modigravi}Fern\'andez-Jambrina, L.; Lazkoz, R. Singular 
 fate of the universe in modified theories of gravity. \emph{Phys. Lett. 
B} \textbf{2009}, \emph{670},  254.

\bibitem{Nojiri:2005sx}
 Nojiri, S.; Odintsov, S.D.; Tsujikawa, S. Properties of singularities
 in (phantom) dark energy universe. \emph{Phys.  Rev.  D}
 \textbf{2005}, {\em 71}, 063004.

\bibitem{IV} D\c abrowski, M.P.D.; Marosek, K. Regularizing 
cosmological singularities by varying physical constants. \emph{JCAP} \textbf{2013} 
\emph{2},  12.

\bibitem{yurov} Yurov, A.V. Brane-like singularities with no brane. \emph{Phys. Lett. B} \textbf{2010}, \emph{689}, 1.

\bibitem{sesto} D\c abrowski, M.P.;  Marosek, K.; Balcerzak, A.
Standard and exotic singularities regularized by varying constants. \emph{Mem. Della Soc. Astron.}  \textbf{2014}, \emph{85}, 44.

\bibitem{grandrip} Fern\'andez-Jambrina, L. Grand rip and grand 
bang/crunch cosmological singularities. \emph{ Phys.  Rev.  D} \textbf{2014}, 
\emph{90},  064014.

\bibitem{ptrsa} Lazkoz, R.; Fern\'andez-Jambrina, L. \emph{Phil. 
Trans. R. Soc. A} Arxiv:gr-qc2111.09068, in press.

\bibitem{Caldwell:2003vq}
 Caldwell, R.R.; Kamionkowski, M.; Weinberg, N.N. 
 Phantom Energy and Cosmic Doomsday. 
 \emph{Phys.  Rev.  Lett.}  \textbf{2003}, {\em 91} 071301.

\bibitem{sudden} Barrow, J.D. Sudden future singularities. \textit{Class. Quant. Grav.}  \textbf{2004}, {\em 21}, 
L79.

\bibitem{odin}
Nojiri, S.; Odintsov, S.D. Quantum escape of sudden future 
singularity. \textit{Phys. Lett. B}  \textbf{2004},
 {\em 595}, 1.
\bibitem{sudden1} Barrow, J.D. More general sudden singularities. 
  \textit{Class.\ Quant.\ Grav.} \textbf{2004}, {\em 21}, 5619.
\bibitem{lake} Lake, K. 
  \textit{Class.\ Quant.\ Grav.}  Sudden future singularities in FLRW 
  cosmologies. \textbf{2004}, {\em 21}, L129.  
\bibitem{odin1} Nojiri, S.; Odintsov, S.D. Final state and 
thermodynamics of a dark energy universe. \textit{Phys.\ 
  Rev.\ D} \textbf{2004}, {\em 70}, 103522. 
\bibitem{mariusz}  D\c abrowski, M.P. Inhomogenized sudden future 
singularities. 
  \textit{Phys.\ Rev.\ D} \textbf{2005}, {\em 71}, 103505.
\bibitem{luis} Chimento, L.P.; Lazkoz, R. On big rip 
singularities. 
\textit{Mod.\ Phys.\ Lett.\ A} \textbf{2004}, {\em 19}, 2479.
\bibitem{mariusz1} D\c abrowski, M.P. Statefinders, higher-order 
energy conditions and sudden future singularities.
\textit{Phys.\ Lett.\ B} \textbf{2005}, {\em 625}, 184. 
\bibitem{barrow2} Barrow, J.D.; Batista, A.B.; Fabris, J.C.; Houndjo, 
S. Quantum particle production at sudden singularities. \emph{Phys. Rev. 
D} \textbf{2008},  \emph{78}, 123508.
\bibitem{barrow3} Barrow, J.D.; Lip, S.Z.W.  Classical stability of 
sudden and big rip singularities. \emph{Phys. 
Rev. D} \textbf{2009}, \emph{80}, 043518.
\bibitem{odin2} Nojiri, S.; Odintsov, S.D. Future evolution and finite-time singularities in 
F(R) gravity unifying inflation and cosmic acceleration. 
\emph{Phys. Rev. D} \textbf{2008}, \emph{78}, 046006.
\bibitem{barrow4} Barrow, J.D.; Cotsakis, S.; Tsokaros, A. A general 
sudden cosmological singularity. \emph{Class. Quant. Grav.}  \textbf{2010},
\emph{27}, 165017.
\bibitem{barrow5} Singh, P. Curvature invariants, geodesics, and the 
strength of singularities in Bianchi-I loop quantum cosmology. \emph{Phys. Rev. D} \textbf{2012}, \emph{85}, 104011.


\bibitem{denk} Denkiewicz, T.;  D\c abrowski, M.P.; Ghodsi, H.; Hendry, M.A.
Cosmological tests of sudden future singularities. \emph{Phys. Rev. D} \textbf{2012},  \emph{85}, 083527.

\bibitem{suddenfirst} Barrow, J.D.; Galloway, G.J.; Tipler, F.J. 
The closed-universe recollapse conjecture. \emph{MNRAS} \textbf{1986}, \emph{223}, 835.

\bibitem{suddenferlaz}  Fern\'andez-Jambrina, L.; Lazkoz, R. 
Geodesic behaviour of sudden future singularities. 
\emph{ Phys. Rev. D} \textbf{2004}, {\em 70}, 121503.
 
 

\bibitem{quiescent} Shtanov, Y.; Sahni, V. New cosmological 
singularities in braneworld models. \emph{Class. Quant. Grav.}
 \textbf{2002}, \emph{19}, L101.


\bibitem{brake} Gorini, V.; Kamenshchik, A.Y.; Moschella, U.; Pasquier, V.
Tachyons, scalar fields, and cosmology. \emph{Phys. Rev. D} \textbf{2004},  {\em 69}, 123512.
 
\bibitem{boost} Barvinsky, A.O.; Deffayet, C.; Kamenshchik, A.Y. 
CFT driven cosmology and the DGP/CFT correspondence. \emph{JCAP}  \textbf{2010}, \emph{5}, 34.


\bibitem{freeze} Bouhmadi-L\'opez, M.; Gonzalez-D\'\i az, P.F.; Mart\'\i n-Moruno, P.
Worse than a big rip? \emph{Phys. Lett.  B} 2008, {\em 659}, 1.
 
\bibitem{tsagas} Barrow, J.D.; Tsagas, C.G. New Isotropic and
Anisotropic Sudden Singularities.  \textit{Class.\ Quant.\ Grav.}
\textbf{2005}, {\em 22}, 1563.

\bibitem{wsing} D\c abrowski, M.P.; Denkiewicz, T. Barotropic index 
w-singularities in cosmology. \emph{Phys. Rev. D} \textbf{2009},
\emph{79}, 063521.

\bibitem{loitering} Shtanov, Y.; Sahni, V. Did the universe loiter at 
high redshifts? \emph{Phys. Rev. D}  \textbf{2005},
 \emph{71}, 084018.

\bibitem{barotrope} Fern\'andez-Jambrina, L. W-cosmological 
singularities. \emph{Phys. Rev. D}  \textbf{2010},
\emph{82}, 124004.

\bibitem{hidden}Fern\'andez-Jambrina, L. Hidden past of dark energy 
cosmological models. \emph{
Phys.\ Lett.\ B} \textbf{2007}, \emph{656}, 9.

\bibitem{initial} Fern\'andez-Jambrina, L. Initial directional 
singularity in inflationary models. \emph{Phys.  Rev.  D} \textbf{2016},
\emph{94}, 024049.

\bibitem{ellis} Ellis, G.F.R.; Schmidt, B.G. Singular space-times. \emph{Gen. Rel. Grav.} \textbf{1977},
\emph{8},  915.
\bibitem{tipler} Tipler, F.J. Singularities in conformally flat 
spacetimes. \emph{Phys. Lett. A} \textbf{1977}, \emph{64}, 8.
\bibitem{krolak} Kr\'olak, A. Towards the proof of the cosmic 
censorship hypothesis. \emph{Class. Quant. Grav.}  \textbf{1986}, \emph{3},
267.
\bibitem{rudnicki} Rudnicki, W.; Budzynski, R.J.; Kondracki, W. 
Generalized strong curvature singularities and weak cosmic
censorship in cosmological space-times. 
\emph{Mod. Phys. Lett. A} \textbf{2006}, {\em 21}, 1501.
\bibitem{puiseux}
 Fern\'andez-Jambrina, L.; Lazkoz, R. 
 Classification of cosmological milestones. 
 \emph{Phys.\ Rev.\  D} \textbf{2006}, {\em 74}, 064030.


\bibitem{graham}Barrow, J.D.; Graham, A.A.H. Singular Inflation. \emph{Phys.  Rev.  D} \textbf{2015}, \emph{91},
083513.  

\bibitem{graham1} Barrow, J.D.; Graham, A.A.H. New Singularities in 
Unexpected Places. \emph{Int. J. Mod. Phys. D} \textbf{2015} 
\emph{24}, 1544012.


\bibitem{fractional} Harigaya, K.; Ibe, M.; Schmitz, K.; 
Yanagida, T.T. 
Chaotic inflation with a fractional power-law potential in
strongly coupled gauge theories.
\emph{Phys.  Lett.  B} \textbf{2013},  \emph{720}, 125--129.

\bibitem{fractional1}  Harigaya, K.; Ibe, M.; Schmitz, K.; 
Yanagida, T.T. Dynamical fractional chaotic inflation. \emph{Phys.  Rev.  D} \textbf{2014},  \emph{90}, 123524.

\bibitem{ellis1} Ellis, G.F.R.; Schmidt, B. 
 Classification of singular space-times. \emph{Gen. Rel. Grav.}
\textbf{1979}, \emph{10}, 989--997.

\bibitem{cosmography} Visser, M. Cosmography: Cosmology without the 
Einstein equations. \emph{Gen. Relativ. Gravit.} \textbf{2005}, \emph{37}, 1541--1548.

\bibitem{liddle} Liddle, A.R.; Lyth, D.H. \textit{Cosmological
Inflation and Large-Scale Structure}; Cambridge University Press:
Cambridge, UK, 2000.

\bibitem{visser} Catto\"en, C.; Visser, M. 
Necessary and sufficient conditions for big bangs, bounces,
crunches,  rips, sudden singularities, and extremality events. 
\emph{Class.\ Quant.\ Grav.} \textbf{2005},  {\em 22}, 4913.

\bibitem{intermediate} Barrow, J.D.; Saich, P. The behaviour of 
intermediate inflationary universes. \emph{Phys. Lett. B}  \textbf{1990},
\emph{249}, 406--410.


\bibitem{cotsakis} Barrow, J.D.; Cotsakis, S.; Trachilis, D. The 
generic sudden singularity in Brans-Dicke theory. \emph{Eur. 
Phys. J. C} \textbf{2020}, \emph{80}, 1197.

\end{thebibliography}
\end{document}